\documentclass[aip,jcp,reprint]{revtex4-2}
\usepackage{graphicx} 
\usepackage{amsmath}
\usepackage{amssymb}
\usepackage{subcaption}
\usepackage{xcolor}
\usepackage{epstopdf}
\usepackage{hyperref}
\usepackage{float}

\begin{document}


\title{Free Energy and Diffusivity in the Fokker-Planck Theory of Polymer Translocation}


\author{Bhavesh R. Sarode}
\author{Harshwardhan H. Katkar}
\email[Author to whom correspondence should be addressed: ]{hkatkar@iitk.ac.in}
\affiliation{FB 457, Department of Chemical Engineering, Indian Institute of Technology Kanpur, Kanpur - 208016, Uttar Pradesh, India}


\date{\today}

\begin{abstract}
We revisit the Fokker-Planck based theory of driven polymer translocation through a narrow nanopore. A bead-spring model of a uniformly charged polyelectrolyte chain translocating through a semi-implicit model of a nanopore embedded in a membrane are used to gain insights into the underlying free energy landscape and kinetics of translocation. The free energy landscape is predicted using metadynamics simulation, an enhanced sampling method. A direct comparison with the theoretical free energy formulation proposed in the literature allows us to introduce a modification related to the entropic contribution in the theory. Additional classical Langevin dynamics simulation runs are performed to obtain the translocation time distribution for polymers of lengths $N$ driven by voltages $V$ through nanopores of radii $r_p$. In agreement with earlier reports, a scaling of the mean translocation time $\langle \tau_\text{LD} \rangle \sim N^\alpha/V$ is observed, with $\alpha \sim 1.40 - 1.48$ depending on the nanopore size. Fitting the mean first passage time given by the Fokker-Planck theory, $\langle \tau_\text{FP}\rangle$,to simulation results helps gain insights into the diffusivity $k_\text{FP}$ used in the theory. We report a scaling of $k_\text{FP}\sim N^\beta$. The $r_p-$dependent values of the exponent $\beta$ significantly deviate from the Rouse theory prediction of $\beta = -1$ for center-of-mass diffusivity of a polymer chain.

\end{abstract}


\maketitle 

\section{Introduction}
\label{sec:intro}
Translocation of polymers through nanopores is encountered in biotechnological applications involving single-molecule detection, particularly in high-throughput DNA sequencing.\cite{Kasianowicz1996, Palyulin2014, Deamer2016} It is a non-equilibrium process in which a single polymer chain passes through a narrow nanopore embedded in a membrane under the influence of a driving force, typically arising from an externally applied trans-membrane voltage.


A parameter of common interest in studying polymer translocation is the mean translocation time $\langle \tau \rangle$ and its dependence on polymer chain length $N$ and externally applied trans-membrane voltage $V$. Experimental investigations into the dynamics of polymer translocation have revealed a complex landscape of scaling behaviors, where $\langle \tau \rangle$ is typically related to $N$ and $V$ as a power law $\tau \sim N^{\alpha}V^{-\delta}$. The reported values for the scaling exponents $\alpha$ and $\delta$ are notably non-universal. For solid-state nanopores for double-stranded DNA (dsDNA), a range of values of $\alpha$ have been reported. Wanunu \textit{et. al.}\cite{Wanunu2008} have reported a crossover behavior, with $\alpha = 1.40$ for short dsDNA molecules (150-3500 bp) which transitions to a much steeper scaling with $\alpha = 2.28$ for longer dsDNA (3500-20000 bp) translocating via SiN$_x$ nanopore of 4 nm diameter. Carson \textit{et. al.} \cite{Carson2014} reported scaling of $\alpha \approx 1.37$ over a wide range of dsDNA lengths (35 to 20,000 bp) in sub-3 nm pores. Storm \textit{et. al.} \cite{Storm2005} reported a scaling exponent of $\alpha \approx 1.27$. While Fologea \textit{et al} \cite{Fologea2007} reported 1.34 in a wider pore for dsDNA (600-20000 bp). Several experiments with solid-state nanopores have confirmed the simple inverse relationship $\tau \sim 1/V$, corresponding to $\delta \approx 1$.\cite{Storm2005, Carson2014, Wanunu2008, Liu2012} In small solid-state nanopores where pore-polymer interactions are strong, an exponential dependence of the translocation time on voltage, $\tau \sim e^{-V/V_0}$, has also been suggested. \cite{Wanunu2008}


Simulation studies have also been used to report the scaling of translocation time with polymer length $N$ and applied voltage $V$. In unbiased translocation (V=0), 2-dimensional (2D) and 3-dimensional (3D) simulations without hydrodynamic interactions (HI) report scaling exponent $\alpha$ ranging from $2.33$ to $2.55$.\cite{Luo2008, Lehtola2010, Polson2014} In the presence of HI, a range of $\alpha = 2.2-2.44$ has been observed.\cite{Gauthier2008, Luo2008, Lehtola2010} For driven translocation in 2D and 3D simulations without HI, the reported scaling exponent ranges from $1.17$ to $1.68$.\cite{Luo2008, Lehtola2008, Vocks2008, Bhattacharya2009, Edmonds2012, Ikonen2012a} However, 3D Langevin dynamics simulations with mesoscale HI reported $\alpha = 1.05 - 1.18$,\cite{Lehtola2009}, while 3D MD simulations including HI have reported a scaling exponent of $1.42$.\cite{Luo2008}



Non-equilibrium theories that model the driven translocation process as out-of-equilibrium explicitly consider the dynamics of the tension propagating along the backbone of the polymer chain as it translocates through the nanopore. \cite{Sakaue2007, Sakaue2010, Rowghanian2011} Brownian dynamics simulations with a time-dependent friction coefficient arising due to the tension propagation have also been performed. \cite{Ikonen2012a, Sarabadani2020} For unbiased translocation, this theory predicts a scaling of $\alpha = {2\nu + 1}$, where $\nu$ is the Flory exponent.

One of the earliest equilibrium theories described unbiased polymer translocation as a diffusive process solely due to thermal fluctuations, where $\langle \tau \rangle$ was calculated from a random-walk model of the polymer chain, resulting in $\alpha = 2$. \cite{Peskin1993} The theory was further developed by incorporating the role of an entropic barrier in determining the translocation time for a polymer chain translocating through a hole (nanopore with length L $\rightarrow 0$).\citep{Sung1996,muthukumar1999} Using self-consistent field theory, Manohar \textit{et al.}\cite{manohar2024} calculated the entropic barrier for a polymer diffusing from one cavity to another through a hole. Muthukumar\citep{Muthukumar2003} developed a Fokker-Planck based formalism for the single-file translocation of a polymer chain through a finite-length nanopore using a 1-dimensional reaction coordinate. The free energy was predicted using entropic contributions from equilibrium polymer chain statistics, an effective pore-polymer interaction term and the contribution due to the externally applied trans-membrane voltage. The formalism invokes the quasi-equilibrium assumption, where the polymer chain is assumed to be equilibrated at faster timescales related to the translocation time. \cite{Chuang2002,Polson2014,Katkar2018b}.

The Fokker-Planck formalism for polymer translocation includes the diffusivity along the 1-dimensional reaction coordinate as a phenomenological parameter. The theory by Muthukumar suggests that this parameter is related to the diffusivity of the center-of-mass of the polymer chain. \cite{Muthukumar2003, Muthukumar2011PolymerTranslocation} Further, this diffusivity is assumed to be uniform throughout the translocation coordinate. Polson and Dunn attributed the diffusivity along the reaction coordinate to the per-segment diffusivity of the polymer chain. \cite{Polson2014} They modeled this diffusivity using the Einstein relation and considered different values of the friction coefficient for polymer segments inside the nanopore versus those outside the nanopore. Dubbledam \textit{et al.} developed a fractional Fokker-Planck model for describing polymer translocation, which introduces a time-dependent diffusivity. \cite{Dubbeldam2011}

In this work, we revisit several aspects of the quasi-equilibrium theory based on Fokker-Planck formalism proposed in Ref. \cite{Muthukumar2003}. The theory relies on two inputs, \textit{viz.} (a) the free energy landscape along the reaction coordinate and (b) the system's diffusivity along the reaction coordinate. The original theory invokes the quasi-equilibrium assumption and further assumes a single-file translocation inside the nanopore to predict the entropic contribution to the free energy at different stages of translocation. Additionally, the system's diffusivity is assumed to be uniform along the reaction coordinate and is treated as a fitting parameter that is independent of $N$. We employ metadynamics, an enhanced sampling method, to estimate the free energy of short polymer chains ($N\in 61-121$) along the reaction coordinate\cite{Bussi2006}. The entropic contribution in the theory is modified based on the results of the metadynamics simulation. We additionally perform 3D Langevin dynamics simulations of long polymer chains (upto $N=1081$) translocating through nanopores of different widths. The results from these simulations provide insight into the diffusivity parameter used in the Fokker-Planck theory. We report a peculiar scaling of the diffusivity parameter with the size of the polymer chain, which in turn depends on the strength of confinement due to the nanopore.

Details of the theory and the simulation methods are described in the following section. Section \ref{sec:results} begins with a discussion about the free energy landscape obtained from metadynamics simulation and the necessary modification introduced in the theoretical free energy. A comparison of the translocation time from Langevin dynamics simulation and the Fokker-Planck theory is also discussed. The scaling of the diffusivity parameter with the size of the polymer is discussed at the end of this section, followed by concluding remarks.

\section{Methods}
\label{sec:methods}
Reduced units are derived for all quantities by choosing a lengthscale of $3$ \AA, a mass scale of $130$ g/mol, and an energy scale of $k_BT$ with $T=300$ K.  Here, $k_B$ is the Boltzmann constant. The derived timescale $t_s = \sigma\sqrt{\frac{m}{\epsilon}} = 2.17$ ps. All quantities reported in the remainder of this manuscript are in reduced units, unless explicitly stated otherwise.
\subsection{Langevin Dynamics Simulation}
\label{sec:methods:LD}
\subsubsection{System Description}
\label{sec:methods:LD:System}
\begin{figure}
    \includegraphics[width=\linewidth]{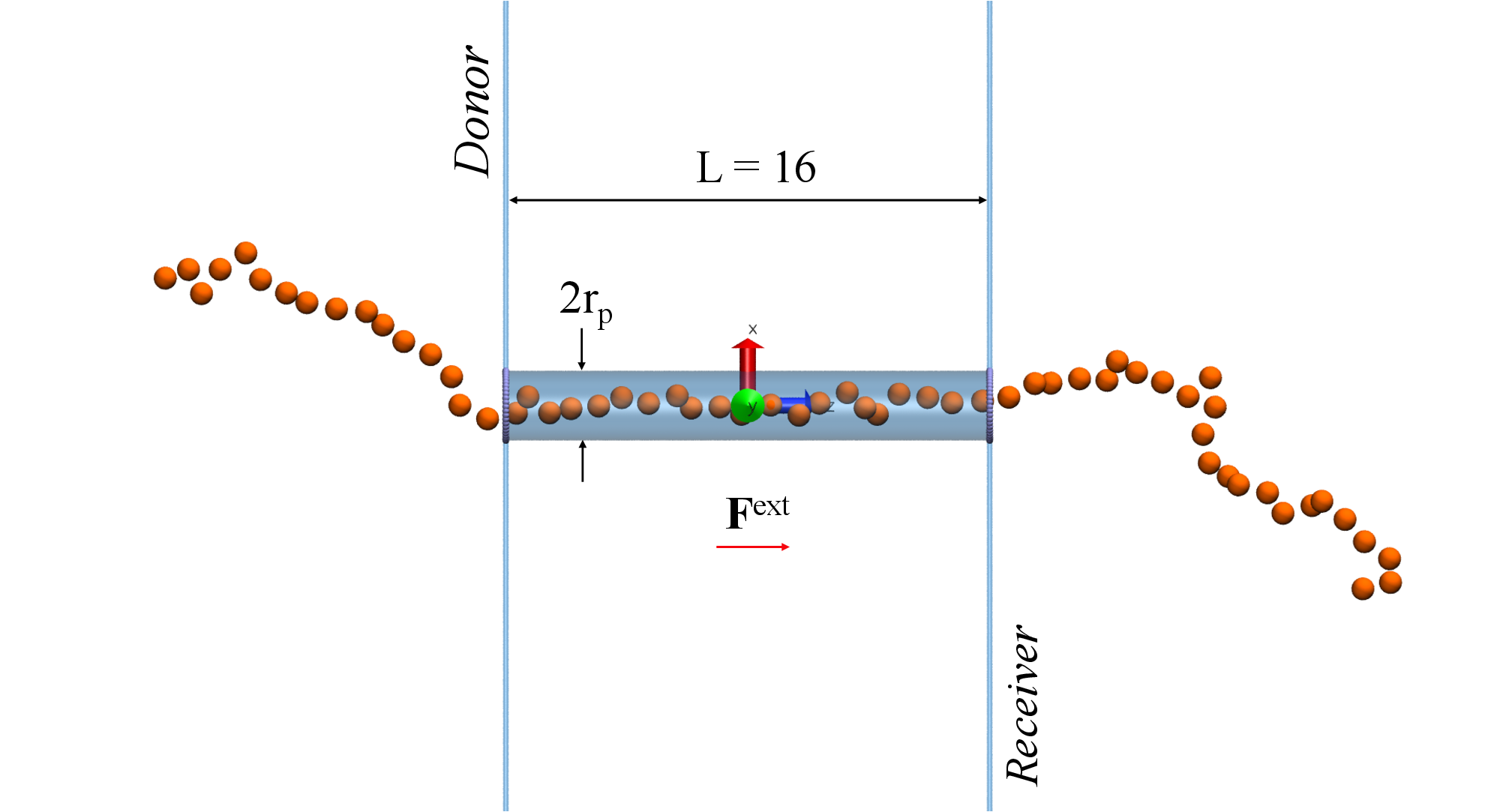}
    \caption{Schematic of the simulation setup. $N$ beads (orange) of the polymer chain translocate through the semi-implicit nanopore of radius $r_p$ and length $L(=16)$ (blue) from donor to receiver compartment in the presence of externally applied force ($\textbf{F}^\text{ext}$).}
    \label{fig:System}
\end{figure}

The 3D system consists of a single linear polymer chain translocating through a cylindrical nanopore of radius $r_p$ embedded in a membrane of thickness $L$ under the action of force, as shown in Figure \ref{fig:System}. The polymer chain is represented using $N$ linearly connected beads. The position vector $\mathbf{r_i}$ of the $i^\text{th}$ polymer bead undergoes dynamics according to the Langevin equations of motion, given by

\begin{align}
    m_i \frac{d^2 \mathbf{r}_{i}}{dt^2} =  - \mathbf{\nabla} (U_{\mathrm{LJ}} + U_b + U_{\mathrm{DH}} ) -\zeta \frac{d \mathbf{r}_{i}}{dt} + \mathbf{F}_{i}^{\text{R}} + \mathbf{F}^{\text{ext}},
    \label{eq:LD1}
\end{align}

where $m_i$ is the mass of the polymer bead and $t$ is time. Conserved forces on the bead are modelled using three potentials.

Excluded volume of the beads is modelled using a Lennard-Jones interaction between two beads $i$ and $j$ at a distance $r$, given by
\begin{align}
    U_{\mathrm{LJ}}(r) =
    \begin{cases} 
    4\epsilon_{\mathrm{LJ}} \left[ \left( \dfrac{\sigma}{r} \right)^{12} - \left( \dfrac{\sigma}{r} \right)^6 \right] + \epsilon_{\mathrm{LJ}}, & \text{for } r \leq 2^{1/6} \sigma, \\
    0, & \text{for } r > 2^{1/6} \sigma.
    \end{cases}
    \label{eq:U_LJ}
\end{align}
The parameters are set to $\epsilon_{\mathrm{LJ}} = 1$ and $\sigma = 1$ for pairs of polymer beads.

For computational efficiency, the membrane with an embedded nanopore is modelled using a semi-implicit representation. An implicit composite ``wall" is created as follows. A cylinder of radius $r_p$ and length $L$ with its axis along the $z$ direction is subtracted from a rectangular block of sides $L_x$, $L_y$, and $L$ using the \textit{fix region} command in LAMMPS\cite{PLIMPTON19951}. The composite wall spans a region from $z=-L/2$ to $z=L/2$ as shown in Figure \ref{fig:System}. The resulting sharp circles at the two ends of the composite wall have ill-defined surface normals, which leads to an unstable simulation. To prevent this, an additional circular ring of radius $r_p$ made of explicit stationary beads is introduced at each of the two ends of the nanopore. Each ring is made of $N_b$ = round($\pi r_p / 0.25$) stationary beads that are uniformly placed along the perimeter of the ring. This composite wall interacts with polymer beads as per equation \ref{eq:U_LJ}, with $r$ representing the distance from particle $i$ to the nearest point on its surface.

For the interaction given by equation \ref{eq:U_LJ} between a polymer bead and either the composite wall or a stationary bead, $\epsilon_{\mathrm{LJ}} = 1$ and $\sigma = 0.625$.

Consecutive polymer beads are connected with a harmonic bond to form a linear polymer chain. The harmonic bond potential between polymer beads $i$ and $i+1$ at a separation distance $r$ is given by the potential
\begin{align*}
    U_b = K(r - r_0)^2,
\end{align*}
where $r_0 = 1$ is the equilibrium bond length and $K = 15480$ is the spring constant. The polymer bead at the center of the chain is neutral, while each remaining bead $i$ carries a charge of $q_i=-e/(\sqrt{4\pi \epsilon_0 \sigma k_BT})=-13.613$ where $\epsilon_0$ is the permittivity of free space, which corresponds to the charge of an electron in reduced units. $e$ is the elementary charge. Debye-H\"{u}ckel potential is used to model the electrostatic interactions between a pair of polymer beads $i$ and $j$ at a distance $r$, as
\begin{align*}
U_{\mathrm{DH}}(r) =
\begin{cases} 
 \frac{q_i q_j}{r} e^{-\kappa r}, & \text{for } r \leq 3\kappa^{-1}, \\
0, & \text{for } r > 3\kappa^{-1}
\end{cases}
\end{align*}
where \(\epsilon_r (= 80)\) is relative permittivity. The inverse of the Debye length,  \(\kappa = 0.3080\), which corresponds to a 0.1 M monovalent salt solution in water. The fluctuation-dissipation theorem governs the relationship between the drag force, characterized by a drag coefficient $\zeta = 1$, and the random force $\mathbf{F}^R$ acting on the monomer due to the implicit solvent. 

A voltage difference of $V$ applied across the nanopore is assumed to result into a uniform electric field of magnitude $E = L/V$ in the $z$-direction inside the nanopore. The resulting external force on a charged polymer bead $i$ located inside the nanopore is given by $\mathbf{F}^{\text{ext}} = q_i \mathbf{E}$, with $\mathbf{E} = (0,0,-E)$. Note that $\mathbf{F}^{\text{ext}} = 0$ for polymer beads outside the nanopore.

\subsubsection{Simulation Protocol}
\label{sec:methods:LD:Simulation}
The system described above is simulated using LAMMPS according to the following protocol. \cite{PLIMPTON19951} The origin is placed at the centre of a periodic box of dimensions $L_x,L_y,L_z$ (Table SI-T19). The centre of the membrane, along with the embedded nanopore of length $L=16$, is at $z=0$, and the nanopore extends from $z=-8$ to $z=8$. The initial conformation of the polymer is such that its first bead is located at $\mathbf{r_1} = (0,0,-7)$ (just inside the nanopore), while positions of the remaining $N-1$ beads are generated randomly, with the distance between consecutive beads equal to the equilibrium bond length ($r_0=1$) and with $z \leq -8$ (to the left of the nanopore). 

To generate a large ensemble of equilibrium chain conformations, we adopted the following computationally efficient protocol. In place of the semi-implicit nanopore, a simple Lennard-Jones wall with $\sigma=0.625$ is set at $z=-8$. The first polymer bead is fixed at its initial location, while the remaining polymer beads undergo Langevin dynamics in the absence of an electric field. The interaction of the first two polymer beads with the Lennard-Jones wall is turned off during the simulation. Initial velocities based on $T=1$ are randomly assigned to all remaining polymer beads. The Langevin equations of motion (\ref{eq:LD1}) are solved numerically using the velocity-Verlet algorithm with a time step $dt = 0.005$. Total simulation time to generate equilibrium chain conformations is ensured to be significantly longer than the Rouse relaxation time of the chain, given as
\begin{align*}
\tau_R = \frac{\zeta  \sigma^2 N^2}{3 \pi^2 k_B T} \approx 3.3774 \times 10^{-2} N^2
\end{align*}

Using this protocol, 2000 statistically independent equilibrium chain conformations are generated for each polymer length $N$. Following this, the Lennard-Jones wall at $z=-8$ used during the equilibration simulation is replaced by the semi-implicit nanopore. All beads of the polymer, including the first bead, underwent Langevin dynamics in the presence of a uniform electric field $\mathbf{E}$ inside the nanopore. A simulation run is considered successful if the entire polymer chain passes through the nanopore and exits from the opposite end. Translocation time $\tau_\text{LD}$ is exclusively determined for successful simulation runs and is defined as the duration between the first time the polymer chain enters the nanopore and the last time it exits the pore. 

The translocation time of polymers of different lengths $N$ translocating through nanopores of different radii $r_p$ under various magnitudes of electric field $E$ is studied. Table \ref{tab:parameters} summarizes different values of these parameters studied in this work. The case of $r_p\rightarrow\infty$ is also simulated following the above protocol, but in the absence of semi-implicit nanopore. 

\begin{table}[]
    \begin{tabular}{| c | l |}
        \hline
        $E$ & 0.01525, 0.01759, 0.0305, 0.04575, 0.061, 0.07652, 0.08794 \\
        \hline
        $r_p$ & 1, 1.4, 1.7, 1.8, 2.1, 2.3, 3.8, 4 \\
        \hline
        $N$ & 61, 81, 101, 121, 241, 361, 721, 1081\\
        \hline
    \end{tabular}
    \caption{Values of parameters used in this work}
    \label{tab:parameters}
\end{table}

\subsection{Fokker-Planck Formalism}
\label{sec:methods:fokker_planck}

The translocation process is modeled using a Fokker-Planck formalism proposed by Muthukumar. \cite{Muthukumar2003} A single reaction coordinate $x$ is used to describe different ``states" of the system as $N$ beads of the polymer chain translocate across the nanopore of length $L$. Of the three stages encountered during the translocation, \textit{viz.} nanopore filling stage, threading stage, and nanopore emptying stage, the reaction coordinate is a measure of the count of polymer beads crossing the entrance of the nanopore during the first two stages. Note that $x=N$, which corresponds to $N$ beads having crossed the nanopore entrance, marks the end of the threading stage. For the nanopore emptying stage, the reaction coordinate measures $N$ plus the length of the empty part of the nanopore. Thus, the reaction coordinate $x\in[0, N+L]$ spans the entire translocation process. The Fokker-Planck equation describes the evolution of the probability $W(x,t)$ of observing the system in state $x$ at time $t$, as

\begin{align*}
\frac{\partial W(x,t)}{\partial t} &= \frac{\partial}{\partial x} \left[J(x,t)\right], & \text{with} \nonumber\\
J(x,t) &= \frac{k_\text{FP} }{k_B T}  \frac{\partial F'(x)}{\partial x} W(x,t)  +  k_\text{FP} \frac{\partial W(x,t)}{\partial x},
\end{align*}
where $F'$ is the dimensional free energy.
Rearranging,
\begin{align}
\frac{\partial W(x,t)}{\partial \tilde{t}} = \frac{\partial W(x,t)}{k_\text{FP}\partial t} &= \frac{\partial}{\partial x} \left[ \frac{\partial F(x)}{\partial x} W(x,t)  +  \frac{\partial W(x,t)}{\partial x}\right], \nonumber\\
\label{eq:non_dimen_FPE}
\end{align}
where $\tilde{t} = k_\text{FP}t$ is the rescaled time and $F = \frac{F'}{k_BT}$ is the dimensionless free energy. We have considered two alternatives for the free energy landscape $F(x)$: one estimated using metadynamics simulation and the other using an analytical theory proposed in Ref.\cite{Muthukumar2003}, with modifications. These are discussed in subsequent sections. The system's ``diffusivity" $k_\text{FP}$ is assumed to be uniform along the reaction coordinate. The net flux $J(x,t)$ consists of drift and diffusion fluxes. The drift arises from the local gradient of the free energy along the reaction coordinate.

The two boundaries are assumed to be absorbing, \textit{i.e.} $W(x=0,t)=W(x=N+L,t)=0$. The initial condition is $W(x,t=0) = \delta(x-x_0)$, where $\delta$ is the Dirac delta function and $x_0$ is the initial reaction coordinate. We use the approximation $\delta(x-x_0) \approx \frac{1}{\sqrt{2\pi}\sigma}\exp({-(x-x_0)/2\sigma^2})$ with $\sigma=0.3$ and with $x_0=2$, the later being consistent with the initial conformation of the polymer chain used in the Langevin dynamics simulation described in Section \ref{sec:methods:LD:Simulation}. 

Equation \ref{eq:non_dimen_FPE} is solved for $W(x,t)$ using the method of lines` approach. The equation is spatially discretized using a finite difference scheme with numerical upwinding, as described in Ref. \cite{Lele1992}. The resulting system of ordinary differential equations is solved using the adaptive time-stepping method implemented in the MATLAB function \textit{ode15s}. \cite{MATLAB} The section SI-3.2 of the supplementary document includes details of the numerical scheme used.

The distribution $g(\tilde{\tau}_\text{FP})$, mean $\langle \tilde{\tau}_\text{FP} \rangle$ and standard deviation $\tilde{\sigma}_{\text{FP}}$ of the rescaled first passage time $\tilde{\tau}_\text{FP} = k_\text{FP}\tau_\text{FP}$ are related to the probability flux at the boundary \( x=N+L \) as \cite{Gardiner, Muthukumar2011PolymerTranslocation}

\begin{align}
g(\tilde{\tau}_\text{FP}) &= \frac{d}{d\tilde{\tau}_\text{FP}} \left(1- \frac{\int_{\tau_\text{FP}}^{\infty} J(N+L,t) \, dt}{\int_{0}^{\infty} J(N+L,t) \, dt} \right) \label{eq:dist_fpt} \\
\langle \tilde{\tau}_\text{FP} \rangle &= k_\text{FP}\langle \tau_{\text{FP}} \rangle = \int_{0}^{\infty} \tilde{t} \, g(\tilde{t})  d\tilde{t}\\
\tilde{\sigma}_{\text{FP}} &= \sqrt{ \left[\int_{0}^{\infty} \tilde{t}^2 \, g(\tilde{t}) d\tilde{t} \right] -  \langle \tilde{\tau}_\text{FP} \rangle^2 }.
\label{eq:sd_fpt}
\end{align}


\subsubsection{Metadynamics Free Energy}
\label{sec:methods:FP:metad}
 The LAMMPS molecular dynamics package \cite{PLIMPTON19951} was patched with the PLUMED version 2.7.6 \cite{BONOMI20091961} is used to estimate the free energy landscape $F_{meta}$ in the absence of an electric field using untempered metadynamics. In metadynamics, a small instantaneous Gaussian bias is slowly added around the instantaneous reaction coordinate $x'(t)$ to penalize revisits to the same location. \cite{Laio2002} We add the instantaneous Gaussian bias intermittently at every 2.5 simulation time units to enhance sampling along the reaction coordinate $x$. Thus, the cumulative bias is
 \begin{equation}
    U_\text{bias}(x,t) = \sum_{t''=2.5k}^{t} h \exp \left( - \frac{(x - x'(t''))^2}{2\sigma_{meta}^2} \right).
 \end{equation}
 We used a Gaussian bias height $h=0.005$ and a Gaussian width $\sigma_\text{meta} = 0.7$. We introduced four additional half-harmonic bias potential walls to limit the sampling of the reaction coordinate within the range of our interest during the metadynamics simulation. The first two are given as $V_l = k_\text{wall}(x-x_{l})^2$ for $x<x_{l}$ and $V_r = k_\text{wall}(x-x_{r})^2$ for $x>x_{r}$. These ensure that the entire polymer chain does not fully escape from the nanopore. We chose $k_\text{wall}=1000$, $x_{l} = 3$ and $x_{r} = (N+L) - 3$. The remaining two bias potentials are $k_\text{wall}(z_1+5)^2$ for $z_1<-5$ and $k_\text{wall}(z_N-5)^2$ for $z_N>5$. These additionally ensure that the first polymer bead does not escape the right end of the nanopore, and the last polymer bead does not escape the left end of the nanopore. The free energy calculation between $x \in [-3+\sigma, N+L-3-\sigma]$ is not significantly affected by these walls.
 
The metadynamics simulation was initialized using one of the equilibrium conformations of the polymer chain as described in section \ref{sec:methods:LD:Simulation}. The reaction coordinate was made continuous and differentiable by defining it as a sum of error functions, as
 \begin{align*}
 x &= {S_1 + S_2}, &\text{where}\\
 S_1 &= \sum_{i=1}^{N-1} \frac{1 + \text{erf}\left(4(z_i + 8) - 2\right)}{2} & \text{and}\\
 S_2 &= \sum_{z'=-8}^8 \frac{1 + \text{erf}\left(4(z_{N} - z') - 2\right)}{2}.
\end{align*}

Here, $S_1$ is a count of all the polymer beads that have crossed the nanopore entrance, while $S_2$ corresponds to the length of the empty part of the nanopore. 

Free energy landscape $F_{\text{meta}}(x)$ is estimated using the cumulative bias added at long time, as
\begin{equation}
\lim_{t \to \infty} U_{\text{bias}}({x}, t) = -F_{\text{meta}}({x}) + C(t)
\end{equation}
where $C(t)$ is the arbitrary uniform energy accumulated until $t$. Five statistically independent metadynamics simulation runs were performed starting from five distinct equilibrated conformations of the polymer chain. Each simulation was run for a sufficiently long time to ensure convergence of the estimated free energy.  Details of the convergence criteria used in our work are discussed in section SI2 of the supplementary document reported as $F_\text{meta}$ in subsequent sections. 

\subsubsection{Theoretical Free Energy}
\label{sec:methods:FP:theoryFE}

Following the theory proposed by Muthukumar\cite{Muthukumar2003}, we write the analytical free energy $F_{\text{th}}(x)$ as a sum of three distinct contributions. The polymer–pore interaction energy $F^{\text{seg}}$ is simply proportional to the number of beads inside the nanopore. The partition sum of a self-avoiding polymer chain of length $N$ with one end adsorbed on a wall is given by Eisenriegler \textit{et al.}\cite{Eisenriegler1982} as $N^{\gamma-1}z^N$, where \(z=6\) is the coordination number of the polymer beads outside the nanopore and the exponent $\gamma = 0.69$ for a self-avoiding chain.\cite{Muthukumar2003} This expression is used to calculate the entropic contribution $F^{\text{ent}}$ associated with the configurational degrees of freedom of the part(s) of the polymer chain outside the nanopore (donor and receiver compartments). However, in contrast to the assumption that a polymer bead loses all configurational degrees of freedom upon entering the nanopore made in the original theory Ref.\cite{Muthukumar2003}, we assume a partial loss of configurational degrees of freedom of a polymer bead upon entering the nanopore. We account for these remaining degrees of freedom in terms of an effective coordination number inside the nanopore, $z_\text{eff}$.  The energy contribution arising from the externally applied voltage across the nanopore is $F^{\text{volt}}$.  During the threading stage, $L$ polymer beads are inside the nanopore, $m = x-L$ polymer beads have translocated into the recipient compartment, and $N-x = N-L-m$ beads remain in the donor compartment. Thus, the three contributions to the free energy during the threading stage are given as

\begin{subequations}
\label{eq:FE_theo2}
\begin{align*}
F^{\text{seg}}(m) &= L\epsilon_\text{FP} 
\\[0.5em]
F^{\text{ent}}(m) &= L \ln\left(\frac{z}{z_\text{eff}}\right) + (1 - \gamma) \ln [m (N - L - m)] 
\\[0.5em]
F^{\text{volt}}(m) &= \frac{q V L}{2}  + \mu m 
\end{align*}
\end{subequations}

where \(\epsilon_\text{FP}\) is the averaged per bead pore-polymer interaction energy inside the nanopore, \(q\) is the charge on each polymer bead, \(V\) is the applied voltage and \(\mu = qV\) is the electrochemical potential difference across the nanopore. Note that in Ref. \cite{Muthukumar2003}, $\epsilon_\text{FP} + \ln{z}$ is treated as a combined per-bead interaction. The complete free energy landscape during all three stages of translocation is given in equations SI-E1a -- SI-E3c of the supplementary document. 

Langevin dynamics simulation of an equilibrated polymer chain conformation in the absence of the nanopore was performed for $\tau_R$ time units after equilibration, with instantaneous ($U_\text{LJ}+U_b+U_{\text{DH}}$) recorded every 50 timesteps to obtain the per-bead average potential energy in the absence of the nanopore. The per-bead average potential energy of the polymer in the presence of the nanopore was determined from those frames of our Langevin dynamics simulation that correspond to the threading stage, \textit{i.e.}, $x\in[L,N]$ at a voltage of 86.7 mV. The per-bead averaged potential energy obtained in the absence of a nanopore is subtracted from the per-bead averaged potential energy in the presence of the nanopore to represent the per-bead pore-polymer average interaction energy \(\epsilon_\text{FP}\). \(\epsilon_\text{FP}\) is found to be nearly independent of $N$, but dependens on $r_p$, as shown in Supplementary Figures SI-F5. Thus, the \(\epsilon_\text{FP}\) is reported as the average value over all polymer lengths $N$ in Figure SI-F6 and in Figure \ref{fig:epsilon_FE_comparision}(a).

The coordination number $z_\text{eff}$ is estimated by minimizing the error given as 
\begin{align}
    \sum_{x_i=L}^{N}\left(F_\text{th}(x_i;z_\text{eff}) - F_\text{meta}(x_i)\right)^2 
    \label{eq:z_eff_minimize}
\end{align}
and averaging over $N$=(61, 81, 101, and 121) for each nanopore size $r_p$. This error is the difference in the energy barrier calculated using metadynamics and that estimated from the theory in the absence of an electric field. Figure SI-F6 shows that $z_\text{eff}$ is independent of N.

\section{Results and Discussion}
\label{sec:results}

\begin{figure}[htbp]
    \centering
    \begin{subfigure}[b]{\linewidth}
        \centering
        \includegraphics[width=\linewidth]{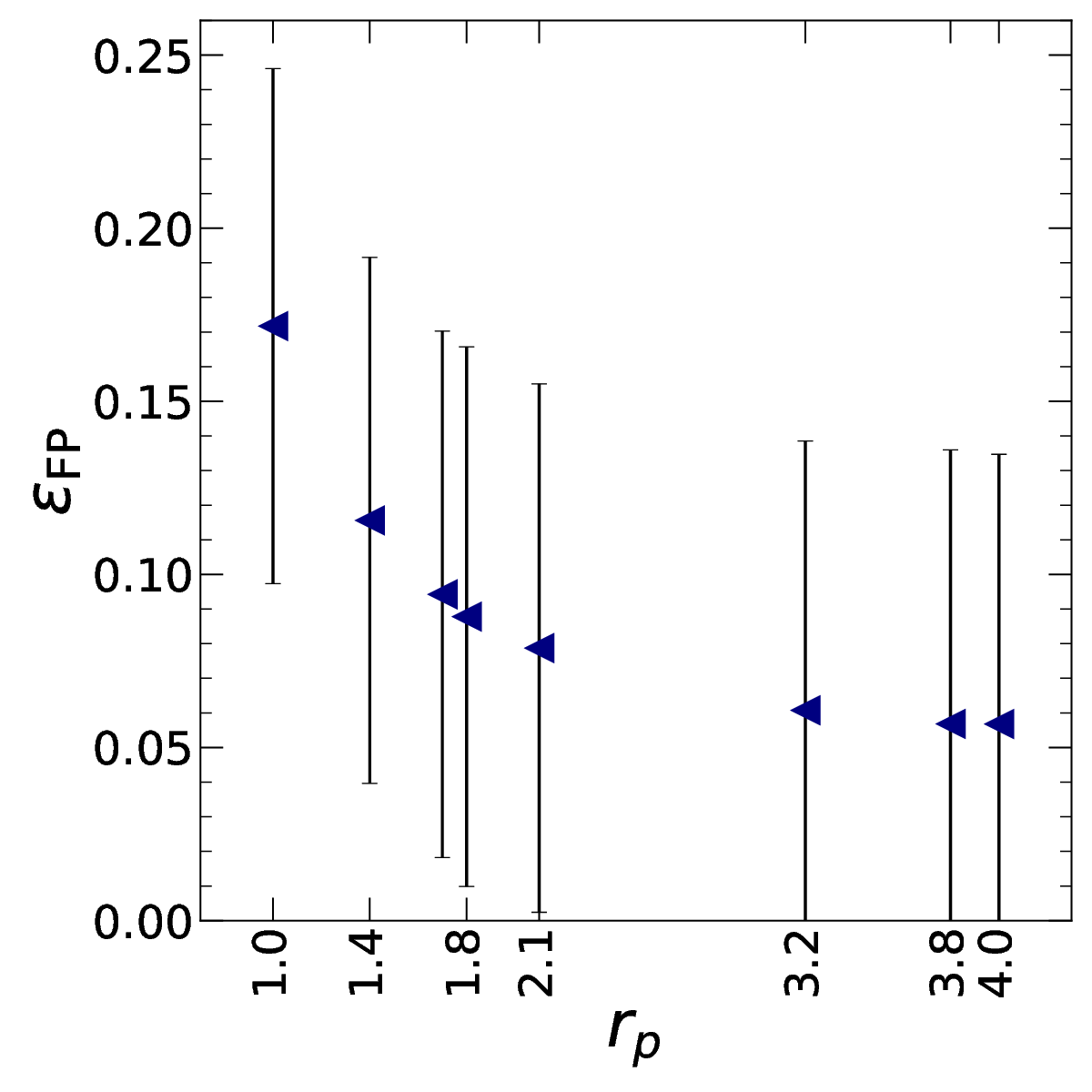}
        \caption{}
    \end{subfigure}
    \begin{subfigure}[b]{\linewidth}
        \centering
        \includegraphics[width=\linewidth]{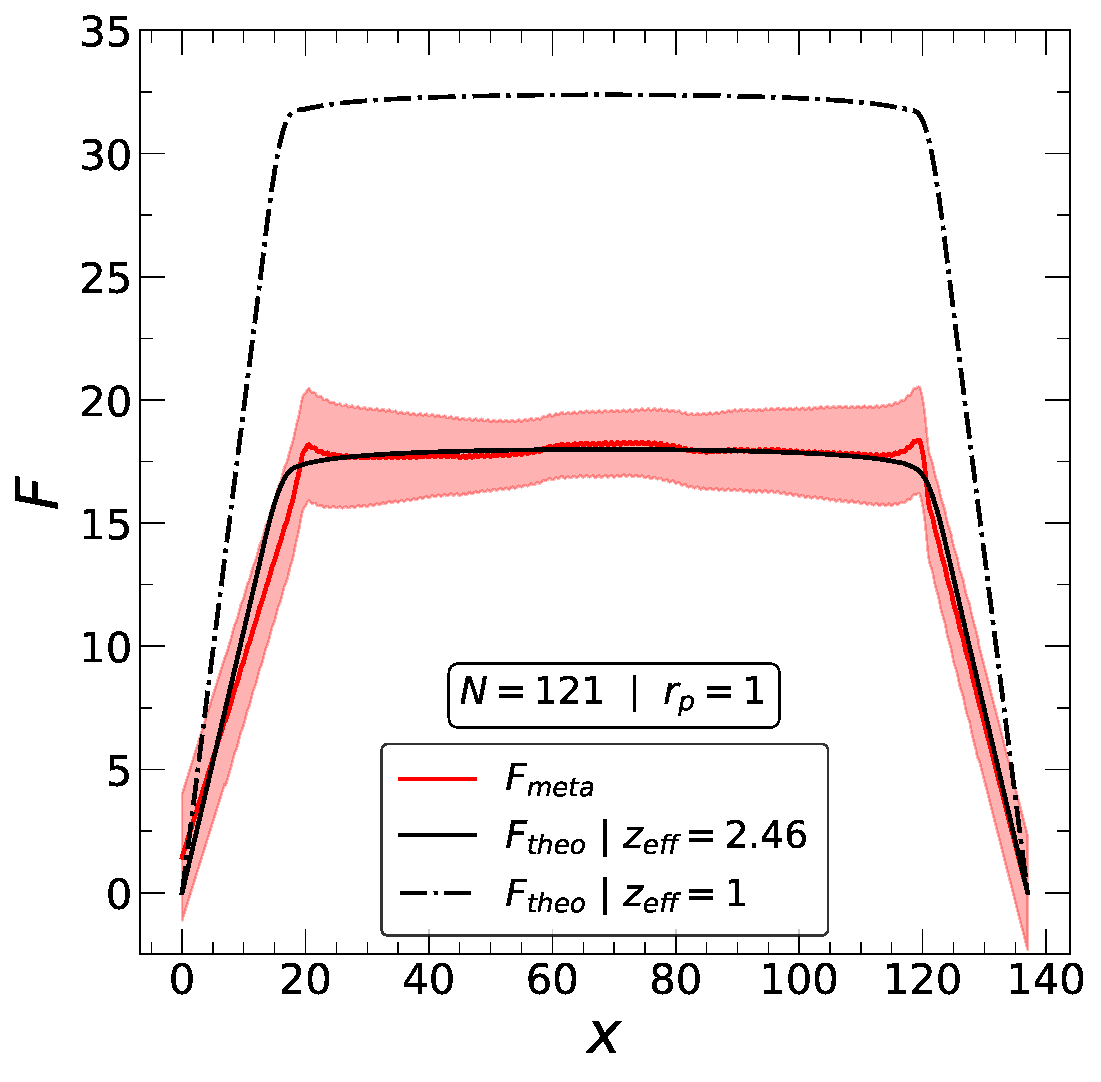}
        \caption{}
    \end{subfigure}
    \caption{(a) Variation of average pore-polymer interactions energy $\epsilon_\text{FP}$ with $r_p$.  
    (b) Comparison of free energy landscapes from theory ($F_\text{th}$) and metadynamics ($F_{\text{meta}}$) for $N=121$ and $r_p=1$. The dot-dashed line shows $F_\text{th}$ for the effective coordination number $z_\text{eff}=1$, while the solid black line shows the same for $z_\text{eff}=2.46$. $\epsilon_\text{FP} = 0.172$ in both.}
    \label{fig:epsilon_FE_comparision}
\end{figure}

The average pore-polymer interaction energy $\epsilon_\text{FP}$ is shown in Figure \ref{fig:epsilon_FE_comparision}(a) for nanopores of different radii $r_p$. $\epsilon_\text{FP}$ is observed to be independent of $N$ and dependent of $r_p$ is shown in Figure SI-F5 and SI-F6(a), respectively. Thus, the average and standard deviation (shown as error bars) in $\epsilon_\text{FP}$ are calculated from its values at different $N$. Polymer beads experience short-range repulsive interactions with the nanopore in our simulation. Stronger confinement in narrower nanopores understandably increases the net repulsion experienced by polymer beads present inside the nanopore. Thus, we observe the largest value of $\epsilon_\text{FP} (= 0.172)$ for our narrowest nanopore with $r_p = 1$. This repulsive interaction is expected to be weaker for wider nanopores. In the limit of $r_p \rightarrow \infty$, we expect $\epsilon_\text{FP} \rightarrow 0$.


The free energy landscape $F_\text{meta}(x)$ estimated using metadynamics simulation for the narrowest nanopore with $r_p = 1$ and for the polymer with $N=121$ in the absence of any trans-membrane voltage difference ($V=0$) is shown in Figure \ref{fig:epsilon_FE_comparision} (b). The shaded region illustrates the uncertainty in our estimate, representing the standard deviation from the five metadynamics simulation replicas. The free energy landscape obtained from the theory $F_\text{th}(x)$ for $V=0$ and $z_\text{eff}=1$ is also shown in the same figure as a black dash-dotted line.  $F_\text{meta}(x)$ qualitatively agrees with $F_\text{th}(x)$ for $z_\text{eff} = 1$ assumed in the original theory. An entropic barrier is predicted by both theory and simulation as the polymer fills the nanopore. However, this barrier is found to be significantly weaker than that predicted by the original theory. The theory assumes that polymer beads lose all entropy and adopt a single conformation when inside the nanopore. In other words, the theory enforces that exactly $L=16$ polymer beads are inside the nanopore in a straight-chain conformation when the nanopore is completely filled during the threading stage. This is strictly true only for the strongest confinement, which corresponds to the case in which the nanopore is just wide enough for the polymer beads to move axially through it without allowing any radial movement, \textit{i.e.} for $r_p \rightarrow \sigma/2 (=0.5$ in our simulation). The nanopores considered in this work are significantly wider than this limit and allow some radial movement of polymer beads inside the nanopore. Figure SI-F3 shows that the average number of polymer beads inside the filled nanopore $N_\text{inside}$ seen in our metadynamics simulation increases with the nanopore radius. More importantly, the number is larger than $L$ even for the narrowest nanopore considered ($r_p = 1$). Thus, we introduce the effective coordination number $z_\text{eff}$ in the theory, and adjust it using equation \ref{eq:z_eff_minimize} to obtain a quantitative agreement between the theoretical and metadynamics free energy landscapes. Figure \ref{fig:epsilon_FE_comparision}(b) also shows the modified theoretical free energy landscape for $z_\text{eff}=2.46$ as a black solid line, which quantitatively agrees with the metadynamics result within error bars.

\begin{figure}[htbp]
    \begin{subfigure}[b]{\linewidth}
        \includegraphics[width=\linewidth]{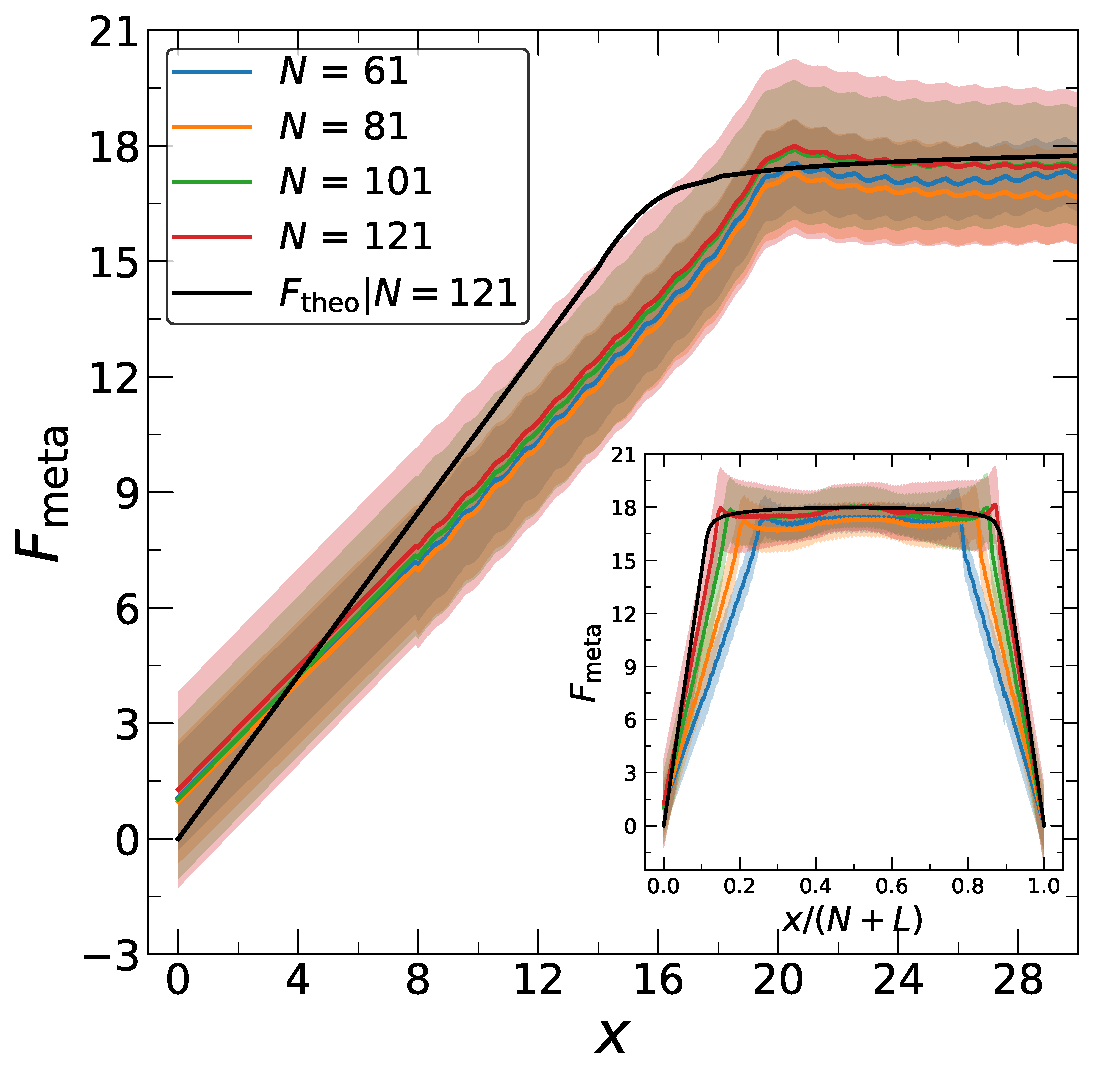}
        \caption{ }
    \end{subfigure}
    \begin{subfigure}[b]{\linewidth}
        \includegraphics[width=\linewidth]{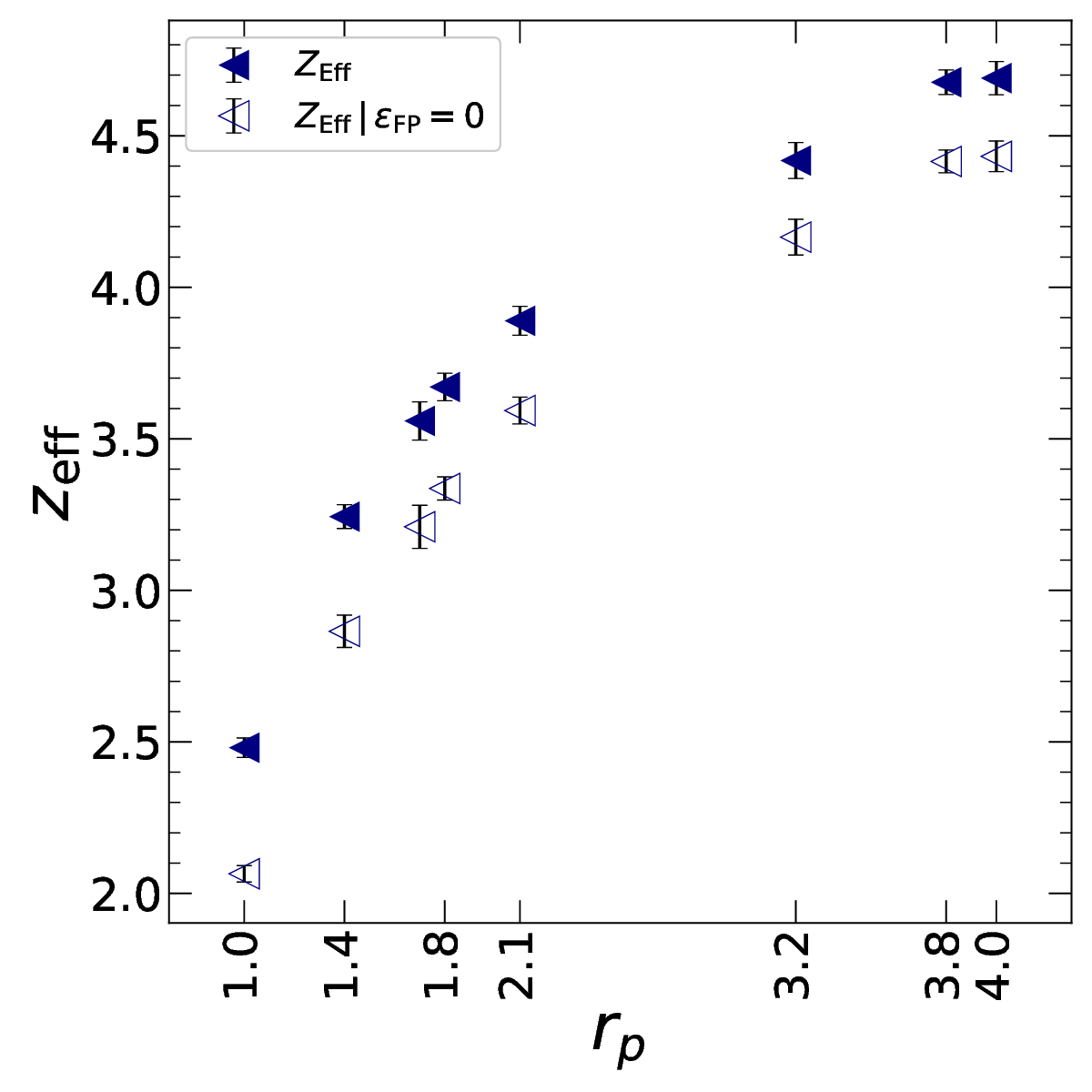}
        \caption{ }
    \end{subfigure}
    \caption{(a) Comparison of $F_{\text{meta}}$ for $N=$ 61, 81, 101, and 121 (colored lines) and $F_{\text{th}}$ (solid black line) for $N=121$ during the nanopore filling stage. Inset shows $F_{\text{meta}}$ and $F_{\text{th}}$ using rescaled reaction coordinates, highlighting the independence of the free energy barrier with respect to $N$. (b) The effective coordination number of polymer beads inside the nanopore $z_\text{eff}$ obtained using $\epsilon_\text{FP}$ values from figure \ref{fig:epsilon_FE_comparision}(a) (filled left-triangles), and using $\epsilon_\text{FP}=0$ (empty left-triangles).}
    \label{fig:FE_wrt_N_Zeff}
\end{figure}

 The effect of polymer length $N$ on the free energy landscapes during the nanopore filling stage for $r_p=1$ and $V=0$ is shown in Figure \ref{fig:FE_wrt_N_Zeff} (a). For the range of polymer lengths used in our metadynamics simulation, the entropic barrier in $F_\text{meta}(x)$ during the nanopore filling stage is found to be nearly independent of $N$. The inset shows the entire free energy landscapes with a rescaled reaction coordinate, $F_\text{meta}(x/(N+L))$. The entropic barrier is observed to be nearly independent of $N$ over the studied range, suggesting that the entropic barrier primarily arises from the loss of entropy of the polymer beads within the nanopore, rather than from those outside the nanopore. $N_\text{inside}$ remains constant for various polymer lengths $N$ in our metadynamics simulations, as shown in supplementary Figure SI-F3. The observed independence of the entropic barrier with respect to $N$ from metadynamics simulation is consistent with the theory, which gives the entropic contribution to the barrier as two terms: a dominant term that scales as the number of polymer beads inside the nanopore ($\approx L\ln(z/z_\text{eff})$), and a weaker term that approximately scales as the logarithm of the polymer length. We expect the entropic barrier to be dependent on $N$ only in the case of exceptionally long polymers translocating through very short nanopores. Thus, for experimentally relevant polymer lengths $N \lesssim O(10000)$ and nanopore lengths $L \gtrsim O(5$ nm), the entropic barrier can be assumed to be nearly independent of $N$.

We note a small discrepancy between the theoretical and metadynamics free energies during the nanopore filling stage. The slope in free energy during the nanopore filling stage for $F_\text{th}$ is steeper than that observed in $F_\text{meta}$. This discrepancy arises because $N_\text{inside}$ is greater in the metadynamics simulation (Figure SI-F3) compared to the assumed value of $L$ in the theory. The discrepancy can be removed by replacing $L$ with $N_\text{inside}$ in the theory (data not shown). However, the value of the mean translocation time does not significantly change when this correction is introduced into the theory. Thus, we do not include this empirical modification to the theory for simplicity.

Wider nanopores result in weaker confinement. Thus, an increase in $r_p$ is expected to decrease the entropic barrier during the nanopore filling stage. Supplementary document Figure SI-F4 shows $F_\text{meta}(x)$ for eight values of $r_p$ keeping $V=0$. As expected, the entropic barrier decreases with an increase in the nanopore radius. As the entropic barrier during the nanopore filling stage is observed to be independent of $N$ in our metadynamics simulation (Figure \ref{fig:FE_wrt_N_Zeff}a) and the effective coordination number inside the nanopore $z_\text{eff}$ is used as a fitting parameter in equation \ref{eq:z_eff_minimize} for matching $F_\text{th}$ with $F_\text{meta}$ at $V=0$, $z_\text{eff}$ is not expected to be a function of $N$ (Figure SI-F6(b)). We report the value of $z_\text{eff}$ averaged over different $N$ as a function of $r_p$ as solid left-triangles in Figure \ref{fig:FE_wrt_N_Zeff} (b).  Error bars show the standard deviation calculated using $z_\text{eff}$ for different values of $N$. $z_\text{eff}$ increases from $\sim2.46$ for $r_p=1$ to $\sim 4.65$ for $r_p=4$, which is consistent with the observation that weaker confinement results into a decrease in the entropic barrier in our metadynamics simulation.

In writing the theoretical free energy, we have written contributions due to the pore-polymer interaction $F^\text{seg}$ and entropy $F^\text{ent}$ as separate terms. However, the combined per-bead interaction term $[\epsilon_\text{FP}+\ln\left(\frac{z}{z_\text{eff}}\right)]$ appears in the resulting free energy $F_\text{th}(x)$. Instead of using $\epsilon_\text{FP}$ calculated from Langevin dynamics simulation (Figure \ref{fig:epsilon_FE_comparision}) (a), one could arbitrarily set $\epsilon_\text{FP}=0$ and obtained a different set of $z_\text{eff}$ values. These are shown as open left-triangles in Figure \ref{fig:FE_wrt_N_Zeff} (b). Thus, either a combination of $[\epsilon_\text{FP}+\ln\left(\frac{z}{z_\text{eff}}\right)]$ with $\epsilon_\text{FP}$ taken from Figure \ref{fig:epsilon_FE_comparision}(a) and $z_\text{eff}$ taken from solid left-triangles in Figure \ref{fig:FE_wrt_N_Zeff}(b) or $\epsilon_\text{FP}=0$ with $z_\text{eff}$ taken from empty left-triangles in Figure \ref{fig:FE_wrt_N_Zeff}(b) results into the same theoretical free energy $F_\text{th}(x)$ which matches the metadynamics free energy. Noting that both these parameters are nearly independent of $N$, but depend on $r_p$, we obtain the theoretical free energy using these parameter values for longer polymers ($N$ = 241, 361, 721, and 1081) beyond those used in our metadynamics simulation.

In experiments, a trans-membrane voltage difference is applied to drive the polymer across the nanopore. The voltage difference results in an additional contribution that helps overcome the strong entropic barrier observed during the nanopore filling stage in the free energy landscape. In theory, this contribution is obtained by adding equations SI-E1c, SI-E2c, and SI-E3c during respective stages of translocation to $F_\text{meta}$ and $F_\text{th}$. The resulting free energy is used in equation \ref{eq:non_dimen_FPE} to compute the rescaled mean first passage time $\langle \tilde{\tau}_{\text{FP}} \rangle$ and its standard deviation from equations \ref{eq:dist_fpt}-\ref{eq:sd_fpt}. As discussed in Section \ref{sec:methods:fokker_planck}, the diffusivity parameter $k_\text{FP}$ is assumed to be uniform along the reaction coordinate $x$. We performed Langevin dynamics simulation in the presence of a trans-membrane voltage difference and compared the computed mean translocation time $\langle\tau_\text{LD}\rangle$ with the rescaled mean first passage time $\tilde{\tau}_{\text{FP}}$ from the theory to gain insights into the parameter $k_\text{FP}$.



\begin{figure}
    \begin{subfigure}[b]{\linewidth}
        \includegraphics[width=\linewidth]{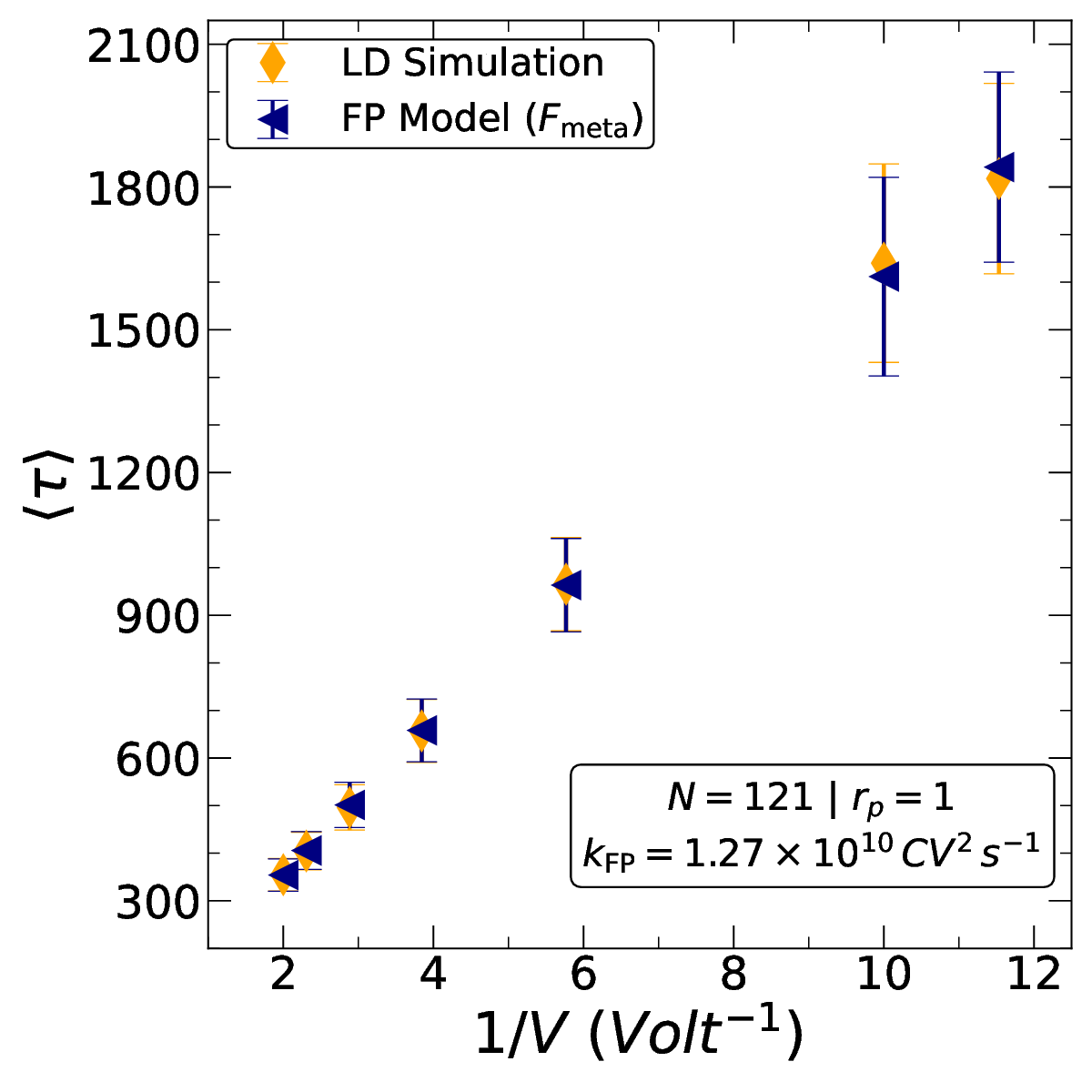}
        \caption{}
        \label{fig:kinetics_comparision_LD_meta}
    \end{subfigure}
    \vspace{0.2cm}
    \begin{subfigure}[b]{\linewidth}
        \includegraphics[width=\linewidth]{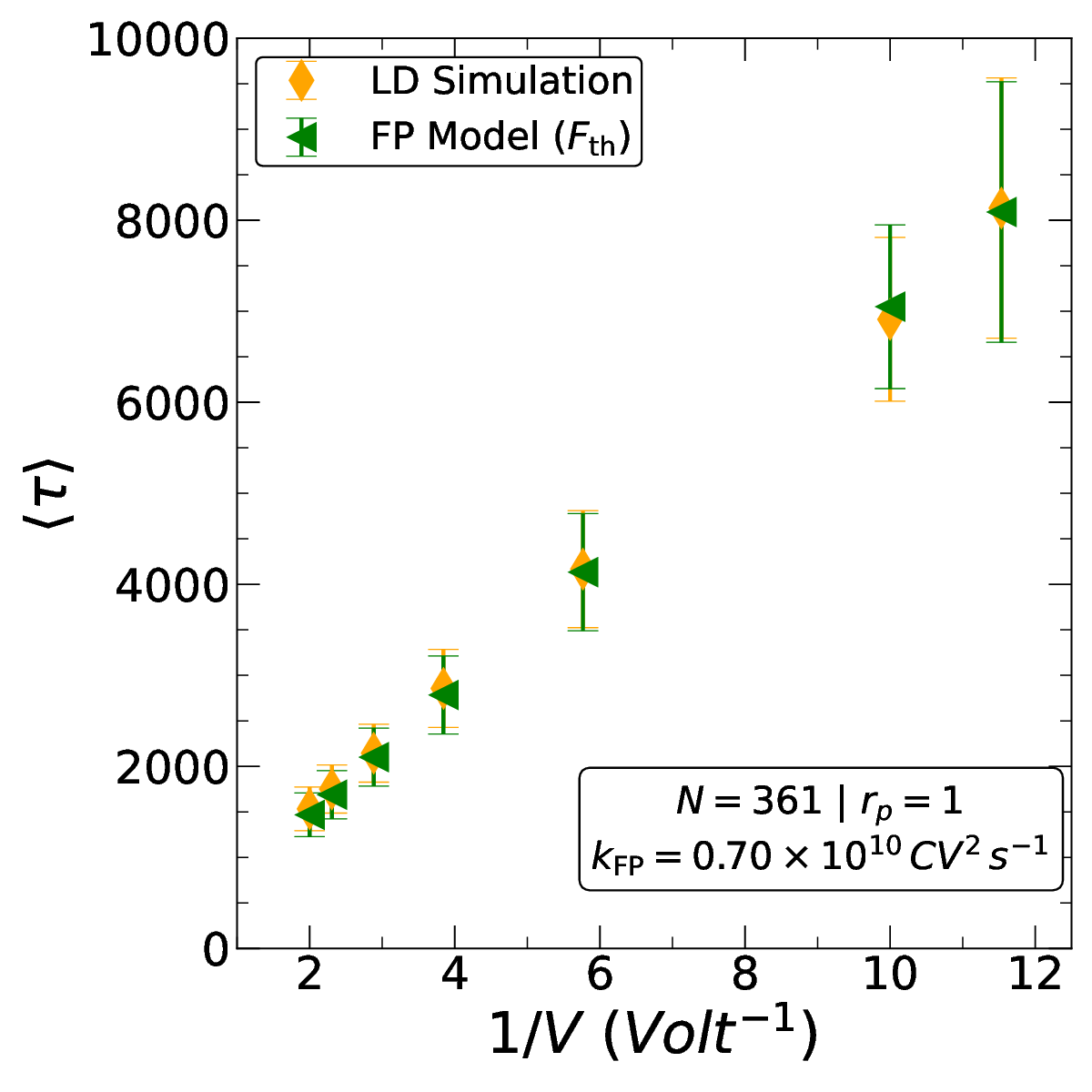}
        \caption{}
        \label{fig:kinetics_comparision_LD_theory}
    \end{subfigure}
    \caption{Variation of mean translocation time from Langevin dynamics simulations \(\langle \tau_{\text{LD}} \rangle\) (yellow diamonds) and mean first passage time from Fokker-Planck formalism with input free energy from (a) metadynamics simulations $\langle \tau_\text{meta} \rangle$ (blue left triangles) and (b) theoretical formulation $\langle \tau_\text{th} \rangle$ (green left triangles), with external trans-membrane voltage $V$.} 
    \label{fig:kinetics_comparision}
\end{figure}

\begin{figure}
    \begin{subfigure}[b]{\linewidth}
        \includegraphics[width=\linewidth]{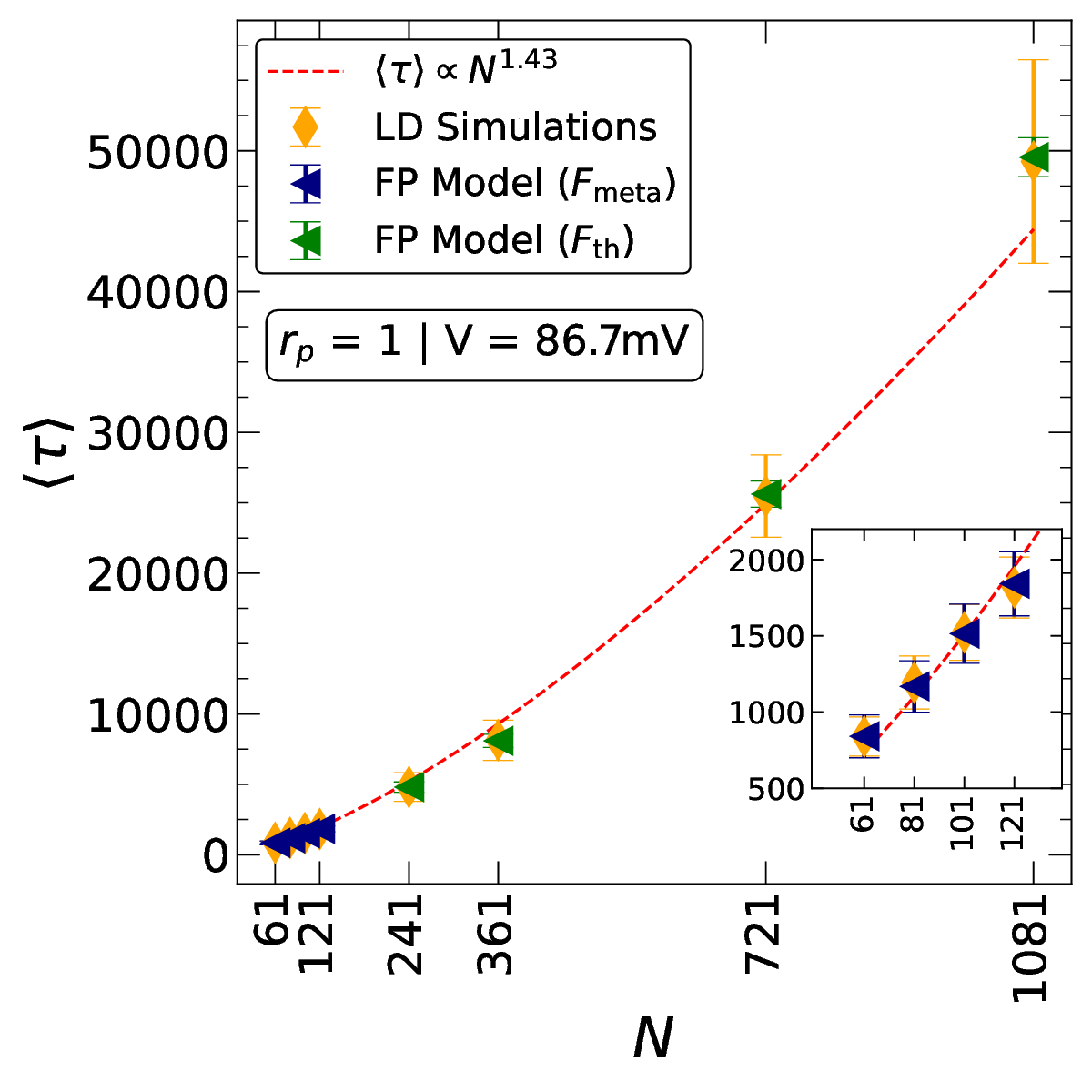}
        \caption{ }
    \end{subfigure}
    \begin{subfigure}[b]{\linewidth}
        \includegraphics[width=\linewidth]{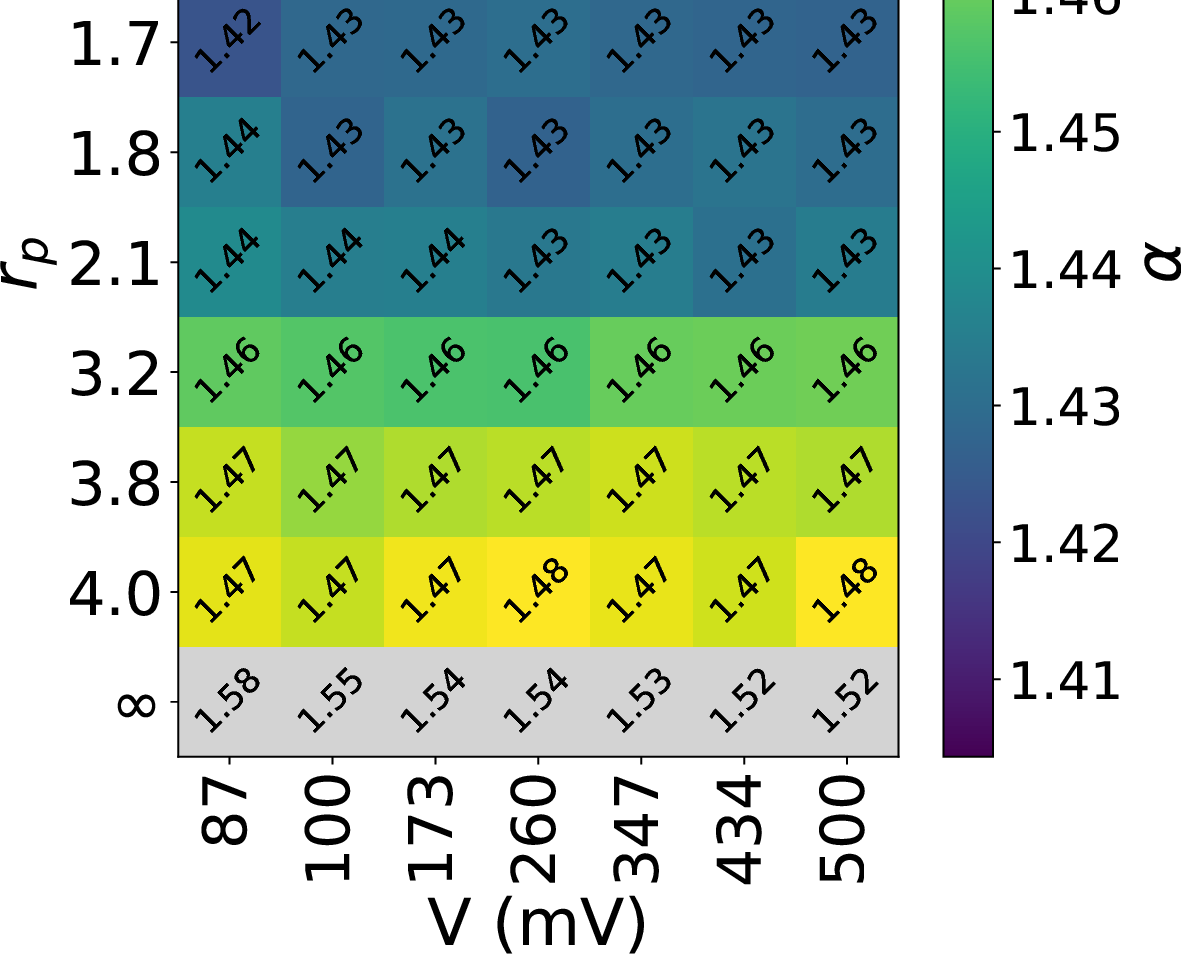}
        \caption{ }
    \end{subfigure}
    \caption{(a) Power law dependence of $\langle \tau \rangle$ on $N$ with the scaling exponent $\alpha=1.43$ for $rp=1$ and $V=86.7$mV. Inset shows a magnified view of the data for smaller $N$. (b) Values of $\alpha$ for different $r_p$ and $V$. $\alpha$ increases with increasing $r_p$, and is independent of $V$. Data for $r_p \rightarrow \infty$ is shown in grey color.} 
    \label{fig:kinetics_comparision_N_rp}
\end{figure}

The translocation time distribution obtained from 2000 Langevin dynamics simulation runs of a polymer chain of length $N=121$ translocating through a nanopore with $r_p=1$ is shown at different trans-membrane voltages $V$ using dashed lines in Figure SI-F9. Figure \ref{fig:kinetics_comparision_LD_meta} shows the corresponding mean translocation time as a function of $1/V$ using diamonds. Error bars indicate the standard deviation of the observed translocation time distribution. The mean translocation time is found to be inversely proportional to voltage. This is in agreement with several experimental \cite{Kasianowicz1996,Matysiak2006,Wanunu2008} and simulation reports \cite{Ikonen2012a,Katkar2018b,Edmonds2012,Vocks2008,Lehtola2008,Bhattacharya2009}. A similar scaling is observed in Figure \ref{fig:kinetics_comparision_LD_theory}, which shows the mean translocation time as a function of $1/V$ for $N=361$ as diamonds. Figures SI-F11(a) to SI-F18(a) show the mean translocation time as a function of $1/V$ for various nanopore radii $r_p$, and for all polymer lengths studied. The scaling of $\langle \tau \rangle \sim 1/V$ is consistently observed for all polymer lengths and all nanopore radii.

The effect of polymer length on the mean translocation time from our Langevin dynamics simulation is shown as diamonds in Figure \ref{fig:kinetics_comparision_N_rp}(a) for $r_p=1$ and $V=86.7$ mV. Over the range of $N$ studied, we observe that for $r_p=1$, $\langle \tau \rangle \sim N^{1.43}$.  A scaling of $\langle \tau \rangle \sim N^{\alpha}$ is observed for all other trans-membrane voltages and nanopore radii studied, with $\alpha \approx 1.40-1.48$ for all nanopore radii except for $r_p\rightarrow\infty$. Figure \ref{fig:kinetics_comparision_N_rp}(b) summarizes the exact values of $\alpha$ obtained for all $V$ and $r_p$. We find a weak but gradual increase in $\alpha$ with increasing $r_p$, but observe no particular trend in $\alpha$ versus $V$ for all finite nanopore radii. At $r_p\rightarrow\infty$, a significant deviation from nearly-single-file translocation for finite nanopores is possible, especially at lower voltages. Although we see a notable decrease in $\alpha$ with increasing $V$ at $r_p\rightarrow\infty$, we do not investigate it further as the mechanism of translocation is likely to be significantly different than that in the case of all finite nanopore radii. Experimentally derived exponents in the range of 1.27-1.40 can be found in the literature, as discussed in Section \ref{sec:intro}. \cite{Storm2005, Wanunu2008, Carson2014}

The Fokker-Planck model is based on conservation of $W(x,t)$, which results in a differential balance of the rate of accumulation of $W(x,t)$ around the ``state" $x$ and the gradient of its flux. The flux is assumed to be arising from the drift and diffusion of the conserved quantity, $W$. The latter is modeled like a Fickian diffusion term with uniform diffusivity $k_\text{FP}$. The former is modeled using a force arising from the local gradient of the free energy, along with a mobility given by the Einstein relation. We have estimated $k_\text{FP}$ by comparing the prediction of translocation kinetics from this theory with Langevin dynamics simulation results.

The parameter $k_\text{FP}$ represents the diffusivity along the reaction coordinate in the Fokker-Planck theory. It is assumed to be independent of the magnitude of the electric field based on the Einstein relation. \(k_{\text{FP}}\) for a given $N$ and $r_p$ was evaluated by minimizing the following sum of squares of errors between the rescaled mean first passage time $\langle\tilde{\tau}_\text{FP}\rangle$ from Fokker-Planck theory and mean translocation time $\langle\tau_\text{LD}\rangle$ obtained from Langevin dynamics simulation for various $V$ used in our Langevin dynamics simulation.
\begin{align}
    \text{error}(k_{FP}) = \sum_{V}^{} \left( \frac{1}{k_{FP}}\langle \tilde{\tau}_\text{FP} \rangle - \langle \tau_{\text{LD}} \rangle \right)^2
    \label{eq:kfp_minimize}
\end{align}
Left triangles in Figure \ref{fig:kinetics_comparision_LD_meta} show the mean first passage time $\langle\tau_\text{FP}\rangle$ for a polymer with $N=121$ translocating through a $r_p=1$ nanopore at various voltages. A single value of $k_\text{FP} = 0.0276$ in reduced units ($=1.27\times 10^{10}$ s${-1}$) results into a remarkable agreement between $\langle\tau_\text{FP}\rangle$ and $\langle\tau_\text{FP}\rangle$ at all voltages. Similarly, $k_{FP} = 0.0152$ results into quantitative agreement between $\langle\tau_\text{FP}\rangle$ and $\langle\tau_\text{FP}\rangle$ at all voltages studied (Figure \ref{fig:kinetics_comparision_LD_theory}).

A comparison of $\langle\tau_\text{LD}\rangle$ and $\langle\tau_\text{FP}\rangle$ as a function of the polymer length is also shown in Figure \ref{fig:kinetics_comparision_N_rp}(a) for $r_p=1$ and $V=86.7$ mV. The inset shows a magnified view for comparison of shorter polymers. It should be noted that the value of $k_\text{FP}$ is different for each $N$ and is obtained by minimizing the error defined in equation \ref{eq:kfp_minimize}.

For a dilute polymer solution in the absence of any hydrodynamic interactions, the Rouse theory predicts the center-of-mass diffusivity of the polymer chain of length $N$ scales as $D \sim N^{-1}$.  Upon adding hydrodynamic interactions, the Zimm theory predicts a scaling of $D\sim N^{-\nu}(\approx N^{-0.588})$. The scaling of $D\sim N^{-1}$ has been reported using Langevin dynamics simulations of a polymer chain diffusing in an infinitely long tube with and without hydrodynamic interactions. \cite{Luo2016, Singh2014, Bhattacharyya2013} In the presence of topological constraints (arising from nearby chains in concentrated solutions), the reptation model predicts that $D\sim N^{-2}$. In the presence of a free energy barrier $F^*$ and in the absence of hydrodynamic interactions, a scaling of $D\sim N^{-1}\exp(-\frac{F^*}{k_BT})$ is predicted. For the Gaussian chain in slit geometry, screening of HI interactions results into $D\sim N^{-0.87}$ for $N \in (100-10000)$.

\begin{figure}
    \begin{subfigure}[b]{\linewidth}
        \includegraphics[width=\linewidth]{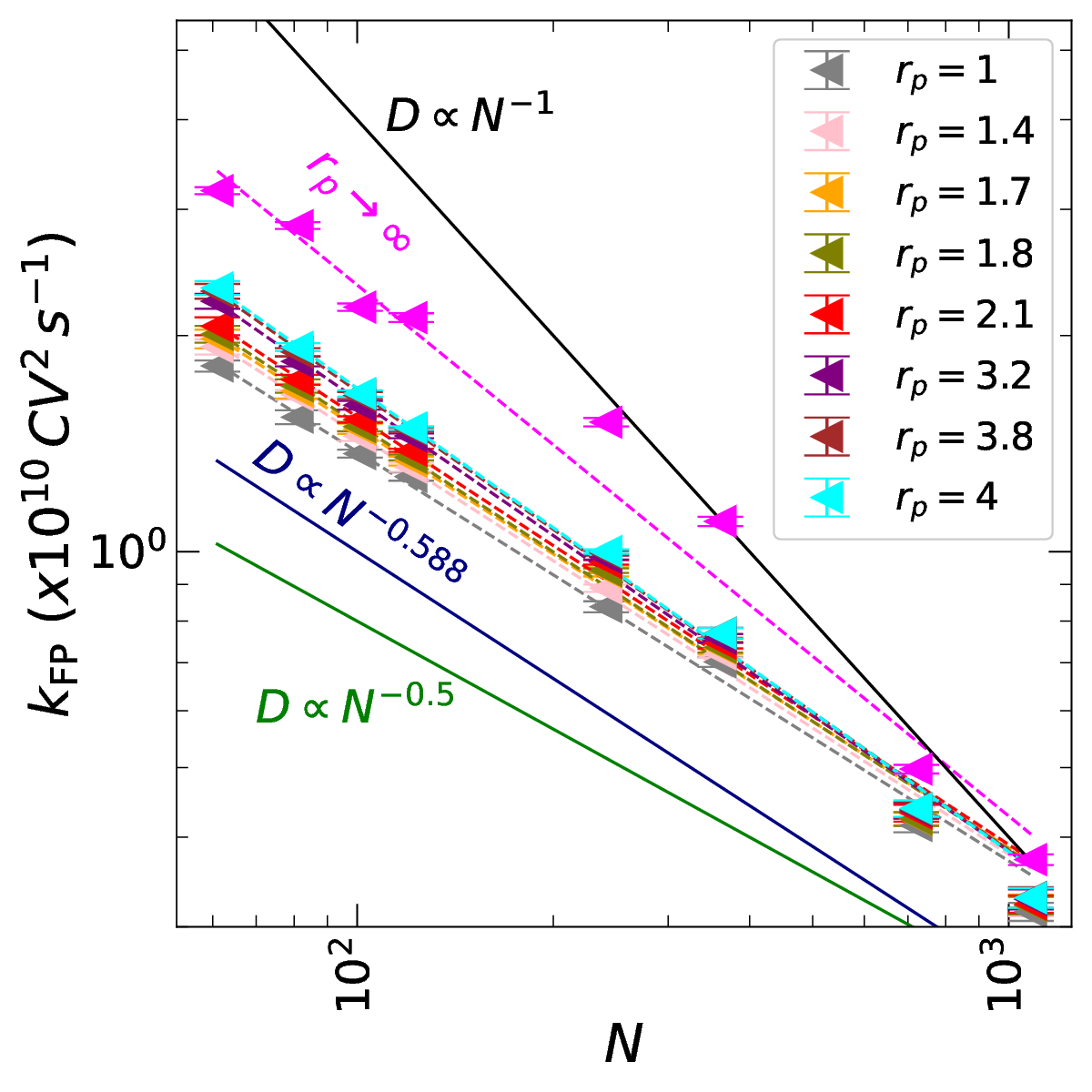}
        \caption{}
        \label{fig:diff_const_wrt_N_loglog}
    \end{subfigure}
    \begin{subfigure}[b]{\linewidth}
        \includegraphics[width=\linewidth]{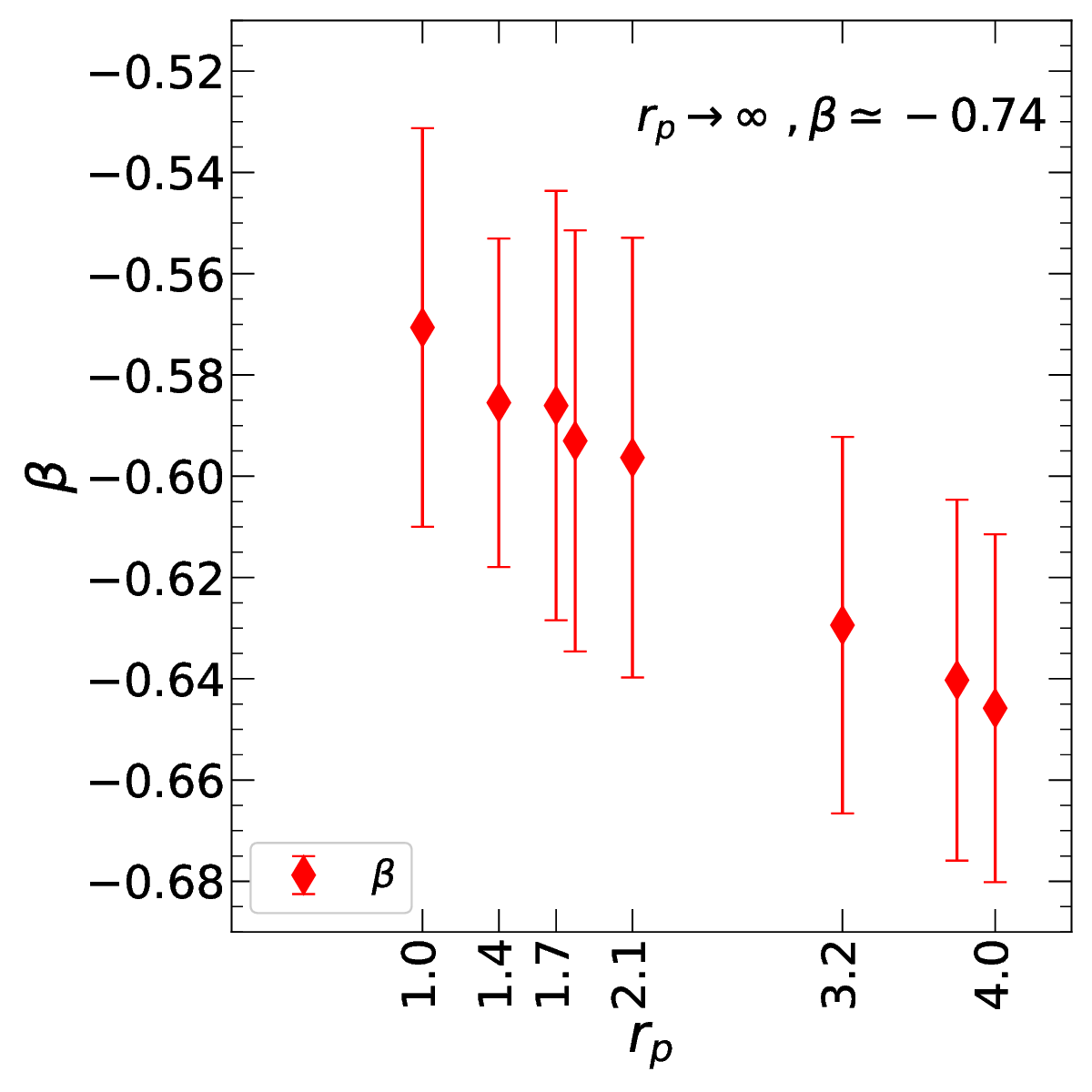}
        \caption{}
        \label{fig:beta_wrt_rp}
    \end{subfigure}
    \caption{(a) A log-log plot of diffusivity $k_\text{FP}$ versus $N$ for different $r_p$, as shown in legend. Magenta left-triangles show the data for $r_p\rightarrow\infty$. Dotted lines show power-law fitting to data corresponding to a given $r_p$. Solid lines show power-law scaling with exponents $-0.5,-0.588$ and $-1$. (b) The power-law exponent $\beta$ for each $r_p$, with error bars showing 95\% confidence interval. The exponent $\beta$ increasingly approaches $-0.5$ with decreasing $r_p$.}
    \label{fig:diffusivity}
\end{figure}

We observe a similar scaling of $k_{FP}\sim N^\beta$ for the diffusivity in the Fokker-Planck equation. The log--log plot in Figure \ref{fig:diff_const_wrt_N_loglog} shows the variation of $k_{\text{FP}}$ with $N$ for different values of $r_p$. The data points for a given nanopore radius fall on a straight line, indicating a power-law behavior. The value of $\beta$ is obtained by minimizing the residual sum of squares of error
\begin{align*}
    \text{error}(\beta) = \sum_{j=1}^{n} \left( k_{\text{FP}, j} - a\, N_{j}^{\beta} \right)^{2}.
\end{align*}

We observe that $\beta \sim -0.57$ for the narrowest nanopore with $r_p=1$, while $\beta\sim -0.74$ for the $rp\rightarrow\infty$ system mimicked by removing the nanopore in our Langevin dynamics simulation. Figure \ref{fig:beta_wrt_rp} shows the values of $\beta$ for different nanopore radii, with error bars indicating the 95\% confidence interval in our fit. We find that $\beta$ increases as $r_p$ decreases. In the limit of $r_p \rightarrow 0$, the exponent approaches $\beta \approx -0.5$.

Although $k_\text{FP}$ is the diffusivity of the probability $W(x,t)$ along the reaction coordinate and not the spatial diffusivity of the center-of-mass of the polymer chain discussed in the preceding paragraph, it has been proposed to represent the polymer chain's diffusivity in the theory \cite{Muthukumar2003, Polson2014}. Our observed scaling is far from the exponent of $-1$ expected for the center-of-mass of a single polymer chain undergoing Rouse dynamics in the absence of any confinement, or from the exponent of $-2$ expected from the reptation theory due to topological constraints. In fact, the observed scaling suggests that the diffusivity of the polymer chain is enhanced despite the confinement it experiences inside the nanopore. This is corroborated by the observed reduction of $\beta$ with increasing $r_p$. For the strongest confinement, $\beta\rightarrow -0.5$, suggesting an enhanced diffusivity with a unique transport mechanism that needs further investigation.

\section{Conclusion}


The Fokker-Planck (FP) formalism is used to gain insights into the kinetics of polymer translocation, where the only input parameters are the free energy and the diffusivity $k_{\text{FP}}$. The free energy landscapes were generated using both the enhanced sampling method $F_\text{meta}$ and the theoretical formulation $F_\text{th}$ for different polymer lengths $N$ and nanopore radii $r_p$. The difference in the entropic barrier in $F_\text{th}$ and $F_\text{meta}$ is adjusted by introducing the effective coordination number $z_\text{eff}$ and the per-segment pore-polymer interaction energy $\epsilon_\text{FP}$ to $F_\text{th}$. These parameters allow us to extend the theory for longer polymers while accurately capturing the effect of nanopore geometry. Our results from LD simulation and FP formalism trans-membrane voltages $V$ show that the mean translocation time $\langle \tau \rangle \sim V^{-1}N^{\alpha}$, where $\alpha$ increases from 1.40 to 1.48 with an increase in $r_p$, which are in good agreement with the experimental observations. The diffusivity of the polymer chain $k_{\text{FP}}$ is calculated by mapping the kinetics from LD simulations to the FP formalism. $k_{\text{FP}}$ varies as $N^{\beta}$ where $\beta$ increases with a decrease in $r_p$, suggesting that diffusivity is enhanced in the presence of confinement. 

\begin{acknowledgments}
This work was supported by the Initiation Grant at Indian Institute of Technology Kanpur. The resources provided by PARAM Sanganak under the National Supercomputing Mission, Government of India at the Indian Institute of Technology, Kanpur are gratefully acknowledged.
\end{acknowledgments}

\bibliographystyle{aipnum4-2}
\bibliography{references}

\end{document}